\documentclass[twocolumn,showpacs,pra,aps,floatfix,superscriptaddress]{revtex4}
\usepackage{bm,multirow,amssymb,amsbsy,amsmath,graphicx,bbold,color}
\makeatletter
\usepackage{pifont}
\makeatother
\newcommand{\tmtextit}[1]{{\itshape{#1}}}

\usepackage{color}
\newcommand{\ket}[1]{\vert #1 \rangle}

\newcommand{\ES}{{\scriptscriptstyle E}}
\newcommand{\SP}{{\scriptscriptstyle S}}
\newcommand{\SE}{{\scriptscriptstyle S\!E}}

\newcommand{\HH}{{\hbox{\small HH}}}

\newcommand{\VV}{{\hbox{\small VV}}}
\newcommand{\ISE}{I_{12}(0)}
\begin{document}
\title{Experimental investigation of initial 
system-environment correlations via trace distance evolution}
\author{Andrea \surname{Smirne}}
\email{andrea.smirne@unimi.it}
\affiliation{Dipartimento di Fisica, Universit{\`a} degli Studi di
Milano, Via Celoria 16, I-20133 Milan, Italy}
\affiliation{INFN, Sezione di Milano, Via Celoria 16, I-20133
Milan, Italy}
\author{Davide \surname{Brivio}}
\email{davide.brivio@unimi.it}
\affiliation{Dipartimento di Fisica, Universit{\`a} degli Studi di
Milano, Via Celoria 16, I-20133 Milan, Italy}
\author{Simone \surname{Cialdi}}
\email{simone.cialdi@mi.infn.it}
\affiliation{Dipartimento di Fisica, Universit{\`a} degli Studi di
Milano, Via Celoria 16, I-20133 Milan, Italy}
\affiliation{INFN, Sezione di Milano, Via Celoria 16, I-20133
Milan, Italy}
\author{Bassano \surname{Vacchini}}
\email{bassano.vacchini@mi.infn.it}
\affiliation{Dipartimento di Fisica, Universit{\`a} degli Studi di
Milano, Via Celoria 16, I-20133 Milan, Italy}
\affiliation{INFN, Sezione di Milano, Via Celoria 16, I-20133
Milan, Italy}
\author{Matteo G. A.  \surname{Paris}}
\email{matteo.paris@fisica.unimi.it}
\affiliation{Dipartimento di Fisica, Universit{\`a} degli Studi di
Milano, Via Celoria 16, I-20133 Milan, Italy}
\affiliation{CNISM, Udr Milano, I-20133 Milan, Italy}
\begin{abstract}
   The trace distance between two states of an open quantum system
   quantifies their distinguishability, and for a fixed environmental
   state can increase above its initial value only in the presence of
   initial system-environment correlations. We provide 
   experimental evidence of such a behavior. In our all-optical
   apparatus we exploit spontaneous parametric down conversion 
   as a source of polarization entangled states, and a spatial light modulator
   to introduce in a general fashion correlations
   between the polarization and the momentum degrees of freedom, which 
   act as environment. 
\end{abstract}
\maketitle
\section{Introduction}
\label{s:intro}
The dynamics of an open quantum system $S$ interacting with an
environment $E$ is usually described by means of a completely positive
trace preserving (CPT) map on the state space of the open system \cite{Breuer2007}. The
very existence of such a map generally requires that the initial
correlations between the open system and the environment can be
neglected, i.e. $\rho_{\SE}(0) = \rho_\SP(0) \otimes \rho_\ES(0)$ where
$\rho_{\SP}(0)=\mbox{Tr}_\ES\left\{\rho_{\SE}(0)\right\}$ and
$\rho_{\ES}(0)=\mbox{Tr}_\SP\left\{\rho_{\SE}(0)\right\}$. However, this
assumption is not always physically justified, especially outside the
weak coupling regime 
\cite{Pechukas1994a-etal}. 
Therefore, different approaches to the description
of the reduced system dynamics in the presence of initial correlations
have been developed in recent years
\cite{Royer1996a-etal, Jordan2004a, Stelmachovic2001a, Aniello2010a, Dijkstra2010a, R2}.
\par
An approach for the study of initial correlations that is based on the use of the trace distance
and that does not rely on the determination of any reduced dynamical map
has been introduced in \cite{Laine2010b}.  
In particular, one can find a clear
signature of initial system-environment correlations as follows: if
the environmental state
is fixed, the trace distance
between any two reduced states can increase over its initial value
only in the presence of initial correlations.
\par
Recently the open system dynamics of two qubits has been
experimentally investigated in all-optical settings, where the system
is represented by the polarization degrees of freedom, and the
environment by the spectral \cite{guo09} or by the momentum
\cite{Cialdi2011a} degrees of freedom.  In the present paper we
provide an experimental proof of the feasibility and effectiveness of
the abovementioned theoretical scheme for the detection of
correlations, observing 
the effect of initial system-environment correlations in the subsequent open system dynamics
by means of the trace distance. In particular, we show an increase of
the trace distance between two reduced states, sharing the same
initial environmental state, over its initial value on both short and
long time scales. Despite the fact that a full tomographic analysis
can be performed, thus showing that the experimental setup can cope
with the most general situation, the growth of the trace distance can
here be detected simply by exploiting visibility data, thus showing that
the theoretical analysis can really lead to efficient detection
schemes.
\par
The paper is structured as follows: In Section \ref{s:td} we briefly present
the general theoretical scheme. In Section \ref{s:exp} we
describe the experimental apparatus in some details, whereas
in Section \ref{s:ic} we describe the use of the spatial light 
modulator to introduce system-environment correlations, and 
analyze the evolution of the trace distance. In Section \ref{s:tomo}
we provide the full tomographic reconsturction of the state under
investigation. Section \ref{s:outro} closes the paper with some
concluding remarks.
\section{Upper bound on trace distance evolution}\label{s:td}
The trace distance between
two quantum states $\rho^1$ and $\rho^2$ is defined as 
\begin{equation}
   \label{eq:2}
   D(\rho^1,
\rho^2) =
\frac{1}{2}\mbox{Tr}\left|\rho^1-\rho^2\right|=\frac{1}{2}\sum_k
|x_k|,
\end{equation}
with $x_k$ eigenvalues of the traceless operator
$\rho^1-\rho^2$, and its
physical meaning lies in the fact that it provides a measure
for the distinguishability between two quantum states
\cite{Fuc99}. 
It is a metric on the space of physical states, so
that for any pair of states $\rho^1$ and $\rho^2$ it holds $0\leq
D(\rho^1, \rho^2) \leq1$.  Every CPT map $\Lambda$ is a contraction
for this metric: $D(\Lambda\rho^1,\Lambda\rho^2)
\leq D(\rho^1,\rho^2)$, a property which will be crucial in the following
analysis.
\par
The dynamics of an open quantum system can be characterized by
investigating the dynamics of the trace distance
between a pair of reduced states $\rho^1_{\SP}(t)$ and
$\rho^2_{\SP}(t)$, which evolve from two different initial total states
$\rho^1_{\SE}(0)$ and $\rho^2_{\SE}(0)$.  The change in the
distinguishability between two reduced states can be interpreted as an
information flow between the open system and the environment
\cite{Breuer2009c-Laine2010a}.
 Indeed, since the
reduced states $\rho^k_{\SP}(t)$ are obtained from the corresponding
initial total states $\rho^k_{\SE}(0)$ $k = 1, 2$ through the composition of a
unitary operation and the partial trace,
the contractivity under CPT maps implies that 
\begin{align}
D(\rho^1_\SP(t),
\rho^2_\SP(t))-D(\rho^1_\SP(0), \rho^2_\SP(0))\leq  \ISE\,,
\end{align}
where
\begin{align}
\ISE \equiv D(\rho^1_{\SE}(0),
\rho^2_{\SE}(0))-D(\rho^1_\SP(0), \rho^2_\SP(0))\,. 
\end{align}
That is, the increase of the trace distance during the time evolution is
bounded from above by the quantity $\ISE$, which represents the
information initially residing outside the open system
\cite{Laine2010b}. It is important to notice that the bound $\ISE$ can
also qualitatively reproduce non-trivial features of the trace
distance dynamics even if it is far from being reached
\cite{Smirne2010c}. If the initial total states are uncorrelated and
with the same environmental state, i.e. $\rho^1_{\SE}(0) =
\rho^1_{\SP}(0)\otimes\rho_{\ES}(0)$ and $\rho^2_{\SE}(0) =
\rho^2_{\SP}(0)\otimes\rho_{\ES}(0)$, then $\ISE = 0$.  Thus, for
identical environmental states, one can find an increase of the trace
distance $$D(\rho^1_\SP(t), \rho^2_\SP(t)) > D(\rho^1_\SP(0),
\rho^2_\SP(0))$$ at a time $t$ only if some correlations are
present in at least one of the two initial total states. 
\section{Experimental setup} \label{s:exp}
In our all-optical experimental setup the total system under
investigation consists in a two-photon state produced by spontaneous
parametric downconversion. We look at the evolution of the two-qubit
polarization entangled state, which represents the reduced system, and
trace out the momentum degrees of freedom, which are not observed and
represent the environment. We exploit a programmable spatial light
modulator (SLM) to impose an arbitrary polarization- and
position-dependent phase-shift to the total state. A linear phase is set
both on signal and idler beams in order to purify the state
\cite{cia10},  whereas an additional, generic, phase function
may be imposed to introduce initial correlations between the
polarization and the momentum degrees of freedom in a very general way.
A further linear phase is then used as a time evolution parameter for
the two-qubit state.
\begin{figure}[h!]
\includegraphics[width=0.62\columnwidth,angle=270]{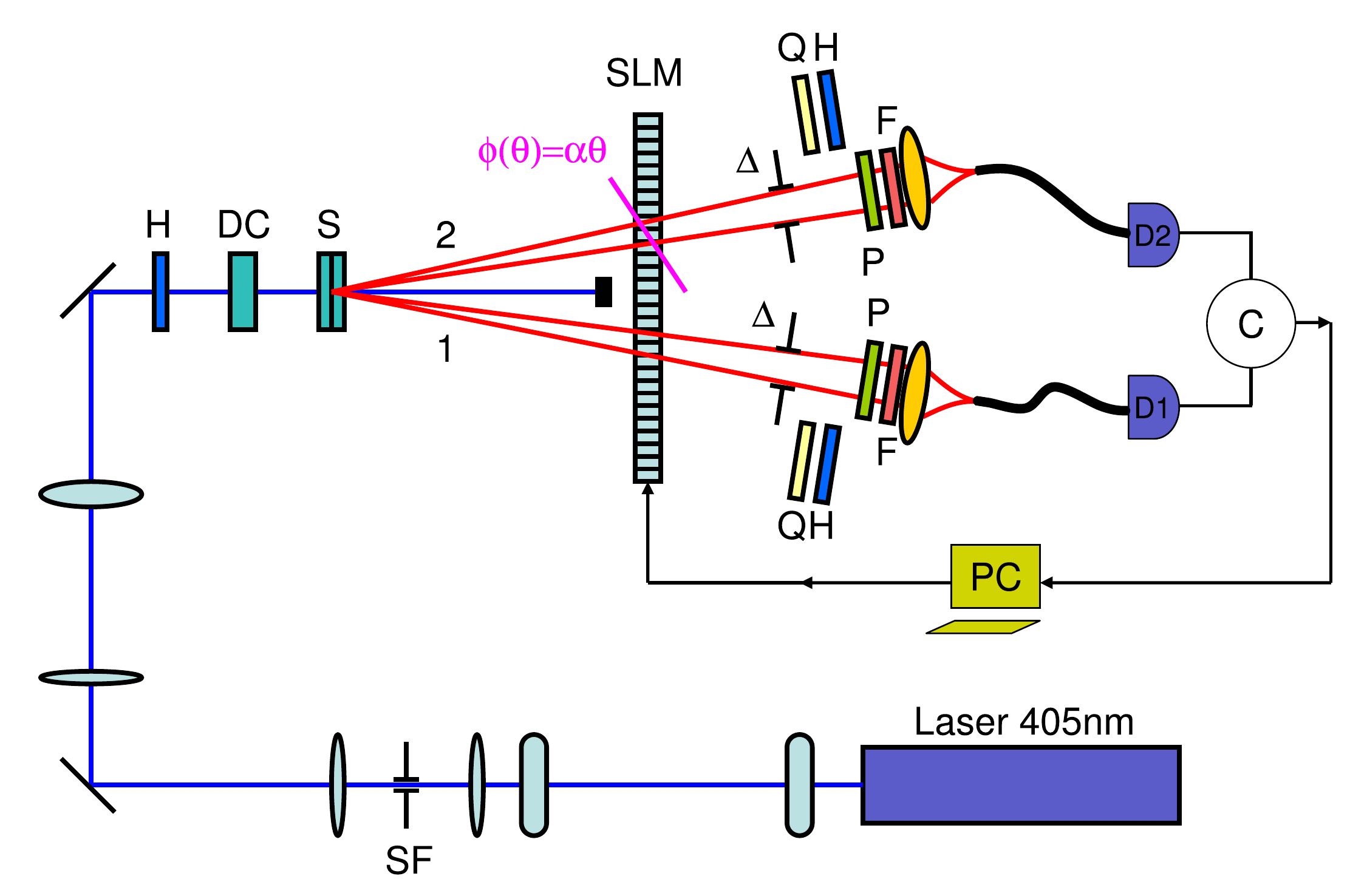}
\caption{(Color online) Diagram of the experimental setup.}\label{setup}
\end{figure} 
\par
The experimental setup is shown in Fig.~\ref{setup}. A linearly
polarized CW, $405\,$nm, diode laser (Newport LQC$405$-40P) passes through
two cylindrical lenses which compensate beam astigmatism, then a spatial
filter (SF) selects a Gaussian spatial profile and a telescopic system
prepares a collimated beam with beam radius of $550\,\mu$m.  A couple of
$1$mm Beta Barium Borate (S) crystals, cut for type-I down conversion,
with optical axis aligned in perpendicular planes, are used as a source
of couples of polarization and momentum entangled photons
\cite{har92,Kwiat99}. The process preserves the total energy and the
transverse momentum. The half wave plate (H) set on the pump path
rotates the pump polarization in order to balance the probability
amplitudes of generating a $\ket{\VV}$ couple of photons in the first
crystal or an $\ket{\HH}$ couple in the second one.  The couples are
generated around a central angle of $\pm 3^{\circ}$ and we select
$\Delta=10\,$mrad with two slits set on signal ($2$) and idler ($1$)
paths.  Two long-pass filter (F) with cut-on wavelength of $780\,$nm set
behind the couplers are used to reduce the background and to select
about $60\,$nm around the central wavelength $810\,$nm,
while the two polarizers (P) are used to perform visibility
measurements as explained later on. The delay time
between the probability amplitudes of generating a $\ket{\VV}$ couple in
the first crystal or a $\ket{\HH}$ couple in the second crystal reduces
the purity of the state.  A nonlinear crystal (DC) with the proper
length and angle is set on the pump path and precompensates this
temporal delay \cite{cia08,kwi09,cia09,bar04,bri08}.  At first order a linear
position dependent phase shift on both channels between $\ket{\HH}$ and
$\ket{\VV}$ photons arises from the angle dependent optical path
followed by $\ket{\VV}$ photons which must traverse the second crystal
\cite{kwi09}.  
\section{System-environment correlations and trace distance evolution}
\label{s:ic}
\subsection{Theoretical description of the experiment}
In our scheme the SLM performs two basic tasks. First, it allows us 
to engineer the initial state by the introduction of an 
arbitrary phase $f(\theta)$. Aside from this, it provides the effective 
system-environment interaction term sensitive to both the polarization
and the momentum degrees of freedom, through the introduction of a
linear phase $\alpha \theta$, where  $\alpha$ is the time evolution
parameter. The total system-environment state is thus given by:
\begin{align}
  | \psi_{\SE}  (\alpha)\rangle & = \frac{1}{\sqrt{2}} \int d \theta d
  \theta' g (\theta) g (\theta')\nonumber\\ 
  &\times \left( |H \theta \rangle |H \theta' \rangle +
  e^{i \left( \alpha \theta + f (\theta)\right)} |V \theta \rangle |V \theta' \rangle \right) . 
  \label{eq:rhose1}
\end{align}
The factorized form $g(\theta)g(\theta')$ is justified by the large
spectral distribution  \cite{Cialdi2011a}. Moreover, $g(\theta)$ is a
Gaussian-like shape function with FWHM of $6$ mrad.  Because of the
phase $f(\theta)$, the state in Eq.~(\ref{eq:rhose1}) is correlated, i.e.
$$\rho_{\SE}(\alpha) = | \psi_{\SE}(\alpha) \rangle \langle
\psi_{\SE}(\alpha) | \neq \rho_\SP(\alpha) \otimes \rho_\ES(\alpha)\,,$$ 
and this is true als for 
the initial total
state, i.e. for $\alpha=0$.  Upon tracing out the momentum degrees of
freedom, the polarization state 
is given by
\begin{eqnarray}
  \rho_{S}(\alpha)  & = & \frac{1}{2}\left(|HH\rangle\langle HH| 
  + \epsilon(\alpha) |VV\rangle\langle HH|\right. \nonumber\\
 && \left.\quad\,+\, \epsilon^{*}(\alpha) |HH\rangle\langle VV| + |VV\rangle\langle VV|\right),
  \label{eq:rhos1}
\end{eqnarray}
where $$\epsilon(\alpha) = \int d \theta |g(\theta)|^2 e^{i
  \left(\alpha \theta + f(\theta)\right)}\,.$$  Since the angular
distribution $g(\theta)$ is symmetric and we use odd functions
$f(\theta)$, the quantity $\epsilon(\alpha)$ is real and 
it equals the interferometric visibility $V(\alpha) =
\mbox{Re}[\epsilon(\alpha)]$.
\par
In order to characterize the effect of the initial system-environment
correlations via the trace distance, we have to monitor the
evolution of two different polarization states obtained from two
different initial total states having the same environmental state. We
compare an initially uncorrelated state $\rho^1_{\SE}(\alpha)$,
corresponding to Eq.~(\ref{eq:rhose1}) for $f(\theta) = 0$, with an
initially correlated state $\rho^2_{\SE}(\alpha)$ for a non-trivial 
function $f(\theta)$. In this way, the reduced
system states $\rho_{\SP}^k(\alpha)$ $k = 1, 2$ are both of the form given by Eq.~(\ref{eq:rhos1}),
with different $\epsilon_k(\alpha)$. Note that the product state
$\rho^1_{\SE}(0)$ differs from $\rho^2_{\SP}(0)\otimes \rho^2_{\ES}(0)$
only for an overall phase term in the integration over $\theta$, 
which has no observable consequences on the
dynamics of the polarization degrees of freedom.  The trace distance
between the two reduced states under investigation is then given by
\begin{align}
\label{eq:td}
D\left(\rho^1_S(\alpha), \rho^2_S(\alpha)\right) &= 
\frac{1}{2}\left|\epsilon_1(\alpha)-\epsilon_2(\alpha) 
\right| \\ 
&= \frac{1}{2}\left| \int d \theta |g(\theta)|^2 
e^{i \alpha \theta} \left(1- e^{i
        f(\theta)}\right)  \right|\,. \nonumber 
\end{align}
Different choices for the initial phase $f(\theta)$ result in 
different dynamical behavior of the trace distance. We 
have exploited this fact to analyze in detail the effect of initial 
system-environment correlations on the subsequent evolution of the open
system.
\subsection{Experimental results}
Experimentally, we have measured the
quantity $\epsilon(\alpha)$ for $f(\theta) = 0$ and $f(\theta) =
\sin(\lambda \theta)$, exploiting its equality with the visibility,
obtained in the standard way by counting the coincidences with polarizers set at
$45^{\circ},45^{\circ}$ and at $45^{\circ},-45^{\circ}$ (see
\cite{cia09} for further details).  The functions of the variable $\theta$ are
discretized by the SLM, and thus become functions of the pixel number
$n$. The resolution is given by $h/D$, where $h=100\,\mu$m is the pixel
width and $D=330\,$mm is the SLM distance from the source. In our
experiment the SLM introduces the functions
\begin{align}
\phi^1(n)&=-a_{opt} (n-n_1)+b \\ 
\phi^2(n,a)&=a_{opt}(n-n_2)+ a (n-n_2)+f(n-n_2)\,,
\notag
\end{align}
on the two
beams respectively, where $a_{opt}=0.1\,$ rad/pixel is an optimal slope
used to achieve the maximal purification of the polarization entangled 
state, and the constant $b$ is used to offset the residual constant
term. The integers $n_1$ and $n_2$ are the central 
pixel numbers on the idler and on the signal beams. The
experimental evolution parameter is then $a=\alpha h/D$ and 
is expressed in rad/pixel. 
\par
\begin{figure}[h!]
\includegraphics[width=0.99\columnwidth]{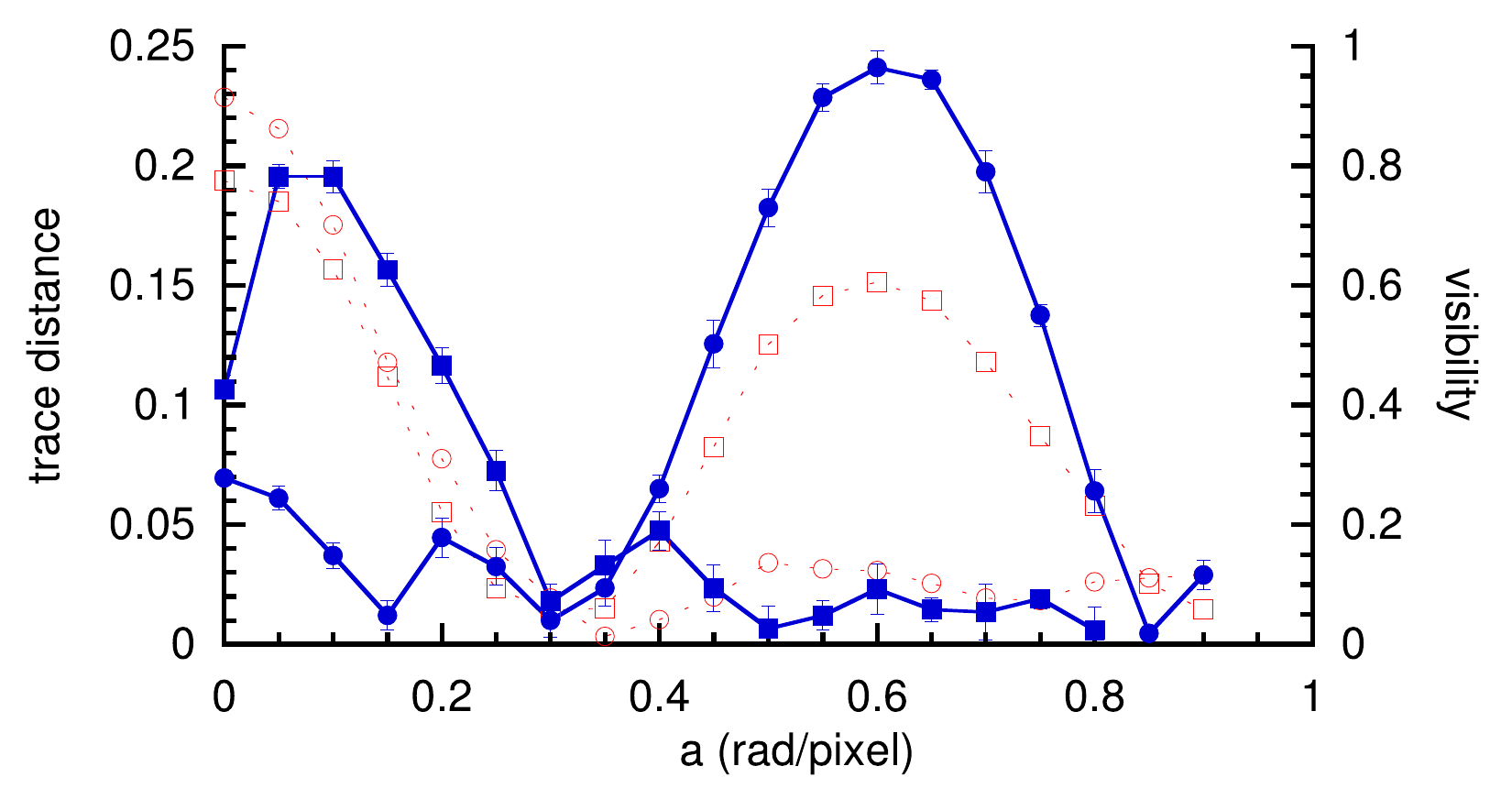}
\caption{(Color online) Trace distance and visibility as a function 
of the experimental evolution parameter $a$, the two quantities are
related through Eq.~(\ref{eq:td}). 
Full circles describe the trace distance between
$\rho^1_S(a)$, i.e. $f(n-n_2)=0$, and $\rho^2_S(a)$ with $f(n-n_2)=
\sin(\lambda (n-n_2))$,  $\lambda = -0.6\,$ rad/pixel. Full squares
describe the trace distance between $\rho^1_S(a)$ and $\rho^2_S(a)$ with
$f(n-n_2)= \tau (n-n_2)$, $\tau = 0.1\,$
rad/pixel. Lines are a guide for the
eye.
Empty circles refer to
visibility with the choice $f(n-n_2)=0$,
whereas empty squares refer to the case in which initial correlations are introduced
through the phase function
$f(n-n_2)=\sin[\lambda (n-n_2)]$.
 For the visibility the uncertainties are within the symbols.
}\label{f2_VTR} \end{figure}
\par
The trace distance is the quantity which reveals
  the presence and the effects of initial correlations, and its
  behavior is reported in Fig.~\ref{f2_VTR}, together with the
  visibility which provides the raw data from which the trace distance
  can be extracted in the present case. In the figure full circles
  describe the trace distance, as a function of the evolution
  parameter $a$, between the reduced state $\rho^1_{\SP}(a)$ evolved
  from the initial total product state, i.e.  $f(n-n_2)=0$, and the
  reduced state $\rho^2_{\SP}(a)$ related to the initial correlated
  state with $f(n-n_2) = \sin (\lambda (n-n_2))$.
The trace distance, after an initial decrease and a first small
oscillation, presents a revival up to a value which is more than three
times the initial one. As expected, the reduced system can access
information which is initially outside it, related to its initial
correlations with the environment. The trace distance reaches its
maximum around $a=0.6$ rad/pixel, toward the end of the monitored time 
interval. The maximum of
the trace distance quantifies the total amount of information which
can be accessed by means of measurements performed on the reduced
system only
\cite{Smirne2010c}. Note that it can be shifted to smaller values of
the evolution parameter $a$ by decreasing the absolute value of
$\lambda$.  Thus, by introducing a sinusoidal phase modulation via the
SLM, we have obtained a behavior of the trace distance which
highlights the presence of initial correlations and their effects in
the subsequent evolution, also for long times \cite{Dajka2010a}.
\par 
The simplest choice for the phase $f(n-n_2)$ in the initially
correlated state $\rho^2_{\SE}(\alpha)$ is a second linear phase
aside from that containing the evolution parameter $a$, i.e. $f(n-n_2) =
\tau (n-n_2)$. Indeed, this corresponds to shift the initially
uncorrelated state $\rho^1_{\SE}(\alpha)$ forward in time by $\tau$.
Then, from the visibility measurement, we can directly obtain
the evolution of the trace distance between $\rho^1_{\SP}(a)$ and
$\rho^2_{\SP}(a)$ with $f(n-n_2)= \tau (n-n_2)$. This is represented by
full squares in Fig.~\ref{f2_VTR}, for $\tau = 0.1\,$ rad/pixel. 
In this case the growth of the distinguishability between
the two reduced states starts from the very beginning of the dynamics.
As expected, the trace distance increases over its initial value,
reaching its maximum value at $a =0.1\,$ rad/pixel and decreasing 
afterwards. The subsequent oscillations can be traced back to
the finite pixel size. Notice also that by using a linear term,  we 
cannot obtain a revival of the trace distance (as in the previous case) 
over its initial value for high values of $a$.
Since in this case $\rho^2_{\SP}(a) = \rho^1_{\SP}(a+\tau)$, the full
squares in Fig.~\ref{f2_VTR} also describe the evolution of the trace
distance between a pair of reduced states occurring at two
different points, separated by $\tau$, of the same dynamics starting
from the initial total product state given by
$\rho^1_{\SE}(0)$. From this point of view, the increase over the
initial value of the trace distance indicates that the single
evolution under investigation is not compatible with a description
through a dynamical semigroup $\Lambda_t$, which could be introduced,
e.g., on the basis of some phenomenological ansatz. Indeed, the
semigroup property $\Lambda_{t+\tau}=\Lambda_t \Lambda_\tau$, together
with the trace distance contractivity under CPT maps, would imply
$D(\rho^1_\SP(t), \rho^2_\SP(t)) = D(\Lambda_t\rho^1_\SP (0),
\Lambda_t\rho^1_\SP(\tau))\leq D(\rho^1_\SP (0),
\rho^1_\SP(\tau))=D(\rho^1_\SP(0), \rho^2_\SP(0))$. However, in
general one cannot discriminate in this way whether the deviations
from the semigroup dynamics are due to correlations in the initial
total state or to other sources of non-Markovianity \cite{R3}.
\section{State reconstruction}\label{s:tomo}
\begin{figure}[h!]
\includegraphics[width=0.44\columnwidth]{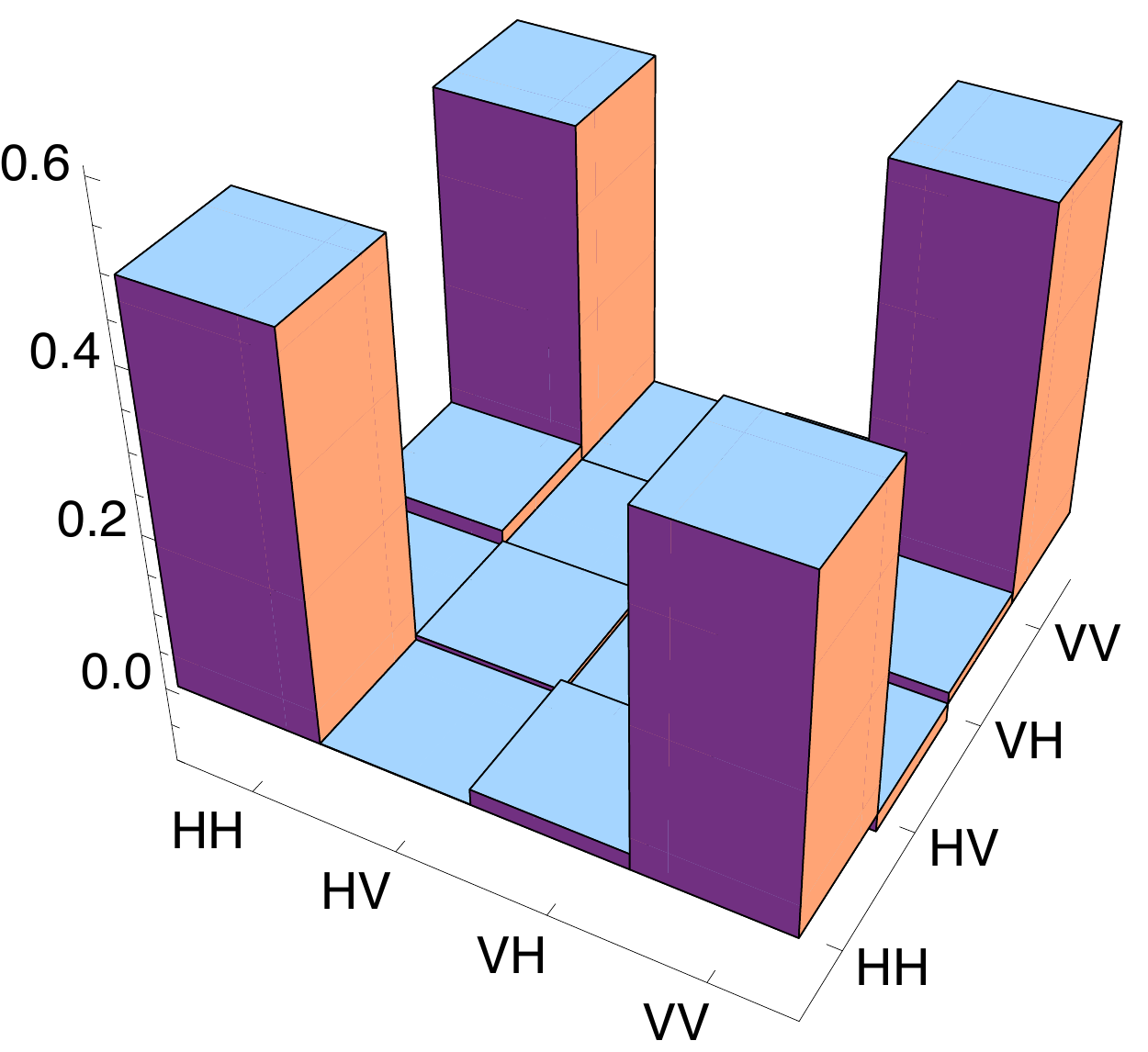}
\includegraphics[width=0.44\columnwidth]{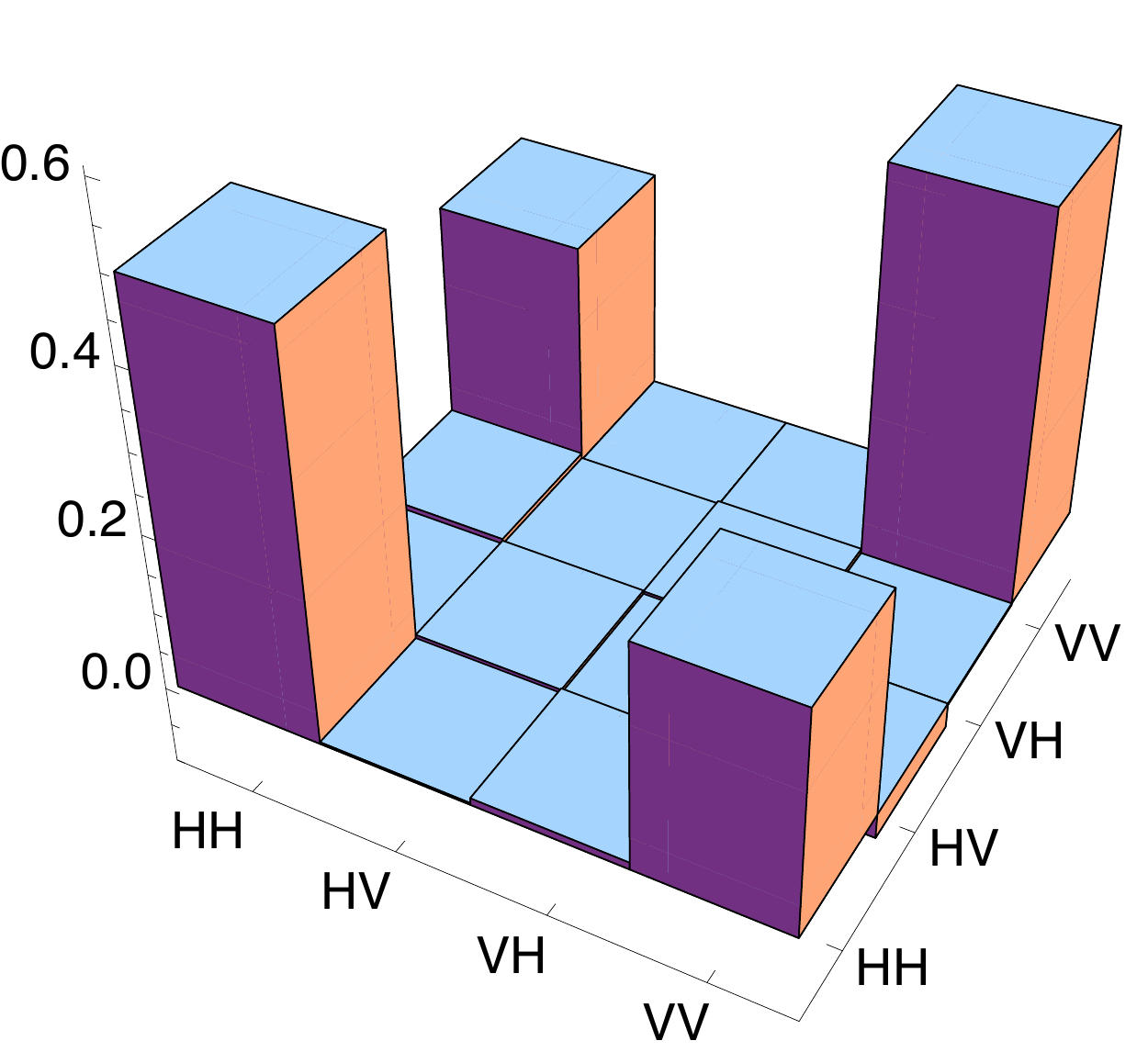}
\caption{(Color online)  Tomographic reconstruction of the two-qubit
density matrix just after the purification (left), without any initial
phase, i.e. for $f(n-n_2) = 0$ and $a=0$. The visibility is $0.914\pm
0.006$.  Tomographic reconstruction for $f(n-n_2)=\sin(\lambda (n-n_2))$ at $a = 0.6$ (right),
i.e. at the maximum of the visibility revival [compare with
Fig.~\ref{f2_VTR}]. The corresponding visibility is $0.605\pm
0.007$}\label{tomographic}\end{figure}

In order to reconstruct the trace distance evolution, we only had to
perform visibility measurements to access the off-diagonal values
$\epsilon_i(\alpha)$. From a mathematical point of view, this
corresponds to explicitly determine the projector operator defining
the trace distance via the relation $D(\rho^1,\rho^2) = \max_{\Pi}
{\rm{Tr}}\left\{\Pi\left(\rho^1-\rho^2\right)\right\}$, 
where the maximum is taken over all the
projectors $\Pi$ or, equivalently, over all the positive operators
${\Pi} \leq \mathbb{1}$. Upon considering the subspace spanned by $\left\{
   |\HH\rangle, |\VV\rangle\right\}$ and the corresponding $\sigma_x$
Pauli matrix, the maximum is here obtained from the projectors on the
eigenvectors of $\sigma_x$, which indeed give back half the difference
between the visibilities. However, in more general situations one
could need a full tomographic reconstruction of the reduced states.
This would be the case in the presence of non-real coefficients
$\epsilon_k(\alpha)$ or when dealing with partially or fully 
unknown states. For this reason, we have also performed state 
reconstruction by polarization qubit tomography. By means of a 
quarter-wave plate, a half-wave plate and a polarizer, we measure 
a suitable set of independent two-qubit projectors
\cite{Banaszek1999a,jam01} 
and then use the maximum-likelihood reconstruction of the 
two-qubit polarization density matrix. In Fig.~\ref{tomographic} 
(left) we show the tomographic
reconstruction of the polarization state 
just after the purification and without any initial correlation, i.e.
for $f(n-n_2) = 0$ and $a=0$. 
The visibility is $0.914\pm 0.006$ (not exactly one 
mostly because of the large spectrum detected). 
In Fig.~\ref{tomographic} (right) we report the two-qubit tomography for
the state characterizing the maximum revival of the visibility in the
presence of initial correlations given by
$f(n-n_2)=\sin[\lambda(n-n_2)]$, i.e. at $a = 0.6\,$ rad/pixel. 
The corresponding visibility is $0.605\pm 0.007$.
\section{Conclusions}
\label{s:outro}
We have reported an experimental observation of the effect of
initial correlations between an open quantum system and its environment
by means of the trace distance.  In particular, we have shown the
increase of the distinguishability between two reduced states, sharing
the same reduced environmental state, over its initial value on both
short and long time scales.
Our all-optical scheme is based on the use of a spatial light modulator, 
which allows us to introduce initial correlations in a very general way.  
In particular, this setup allows to engineer different kinds of dynamical 
behavior of the trace distance, so that one can, e.g., tune the position 
and the amplitude of the revival points of the distinguishability.
\par
\emph{Note added in proof.} Recently, we became
aware of \cite{guo11}, where initial correlations between the polarization
and the spectral degrees of freedom of single photon states are experimentally
witnessed by means of the trace distance.
\begin{acknowledgments}
A.S. and B.V. acknowledge financial support by MIUR,
under PRIN2008. The authors thank H.-P. Breuer, E.-M.~Laine, S. Maniscalco and J. Piilo for 
useful discussions.
\end{acknowledgments}

\end{document}